\documentclass{IEEEtran}
\usepackage[latin9]{inputenc}
\usepackage{amsmath}
\usepackage{amssymb}
\usepackage{graphicx}

\makeatletter
\usepackage{cite}
\usepackage{amsfonts}\usepackage{graphicx}
\usepackage{textcomp}
\usepackage{nicefrac}\def\BibTeX{{\rm B\kern-.05em{\sc i\kern-.025em b}\kern-.08em
T\kern-.1667em\lower.7ex\hbox{E}\kern-.125emX}}
\makeatother

\begin{document}
\title{Integrated real-time supervisory management for off-normal-event handling
and feedback control of tokamak plasmas}
\author{Trang VU$^{1}$, Federico Felici$^{1}$, Cristian Galperti$^{1}$,
Marc Maraschek$^{2}$, Alessandro Pau$^{1}$, Natale Rispoli$^{3}$,
Olivier Sauter$^{1}$, Bernhard Sieglin$^{2}$, the TCV Team$^{4}$
and the MST1 Team$^{5}$ \thanks{\emph{$^{1}$} \emph{Ecole Polytechnique Fédérale de Lausanne (EPFL),
Swiss Plasma Center (SPC), CH-1015 Lausanne, Switzerland} (e-mail:
author@epfl.ch), $^{2}$ Max-Planck-Institut for Plasmaphysics, D-85748
Garching, Germany, $^{3}$ Institute for Plasma Science and Technology,
CNR, via Cozzi 53, 20125 Milan, Italy, $^{4}$ See author list of
S. Coda et al 2019 Nucl. Fusion 59 112023, $^{5}$ See author list
of B. Labit et al 2019 Nucl. Fusion 59 086020.}}
\maketitle
\begin{abstract}
For long-pulse tokamaks, one of the main challenges in control strategy
is to simultaneously reach multiple control objectives and to robustly
handle in real-time (RT) unexpected events (off-normal-events - ONEs)
with a limited set of actuators. We have developed in our previous
work (cf. \cite{TrangVU19_FED}) a generic architecture of the plasma
control system (PCS) including a supervisor and an actuator manager
to deal with these issues. We present in this paper recent developments
of real-time decision-making by the supervisor to switch between different
control scenarios (normal, backup, shutdown, disruption mitigation,
etc.) during the discharge, based on off-normal-event states. We first
standardize the evaluation of ONEs and thereby simplify significantly
the supervisor decision logic, as well as facilitate the modifications
and extensions of ONE states in the future. The whole PCS has been
implemented on the TCV tokamak, applied to disruption avoidance with
density limit experiments, demonstrating the excellent capabilities
of the new RT integrated strategy. 
\end{abstract}

\begin{IEEEkeywords}
tokamak plasma control system, supervisory control, integrated control,
off-normal-event handling 
\end{IEEEkeywords}

\section{Introduction}

\label{sec:introduction} \IEEEPARstart{T}{he} development of an
advanced tokamak plasma control system (PCS) has recently gained more
attention with the requirements of a robust off-normal-event (ONE
- plasma or subsystem/plant failures) handling and of an integrated
control approach. These are crucial to ensure a feasible discharge
both for the plasma and the plasma-facing components in long-pulse
tokamaks like ITER. On the one hand, this advanced PCS will act as
the first line of defense of disruption, where all the plasma energy
is released in few ms, to avoid unnecessary mitigation actions. On
the other hand, it must be able to reach the desired discharge performance
by simultaneously fulfilling multiple control tasks (control objectives)
with a minimal set of actuators and diagnostics. Our work improves
the entire chain of the tokamak-agnostic layer in the PCS including
a ONE monitor, a supervisor, an actuator manager and controllers,
as well as demonstrates the efficiency of the proposed approach via
the first applications on TCV (cf. \cite{TrangVU19_FED}).

Regarding ONE handling, several work has focused on the use of a discharge
manager to decide appropriate actions depending on the seriousness
of the events (\cite{Nouailletas13,Eidietis_NF18,Maraschek18}). In
this work, different ONEs categories are distinguished and several
control scenarios are investigated as well. The basic idea of ONE
handling is shown in \cite{Nouailletas13}, where a discharge management
system plays the roles of both a ONE monitor to classify the events
and a supervisor to select a control scenario. In \cite{Eidietis_NF18},
a supervisory logic, using finite-state machines\footnote{a representation of an event-driven (reactive) system which can be
in one of a finite number of states depending on its previous condition
and on the present values of its inputs.}, is developed for event detection. In this early stage, the simple
threshold test on the individual event can only trigger a soft-shutdown
or a mitigation scenario. In \cite{Maraschek18}, based on disruption
root causes, different decentralized handlers are deployed with their
pre-assigned actuators to directly tackle ONEs. 

However, in these works, the supervisory decision is done by the selection
of reactions as well as the corresponding actuators via the prioritization
of ONEs. This leads to a direct link between ONEs and actuators. In
other words, the supervisor needs to be aware of ONE nature and tokamak
specific actuator systems. Here, we propose a systematic way to handle
ONEs by the supervisor. Therefore, more stages are necessary to clearly
classify the \emph{danger level, }the \emph{reaction level} for each
ONE, and a \emph{ONE reaction to scenario} (OS) mapping is finally
used to allow the supervisor to automatically switch between different
control scenarios (normal, backup or shutdown scenarios, etc.). Since
the supervisor only takes care of selecting an appropriate scenario,
the actions to deal with ONEs, once they are detected, are (flexibly)
customized as a list of prioritized control tasks in different control
scenarios. This leads to an automatic actuator resource assignment
of the actuator manager and control (feedback) actions of the controllers
(see Fig. \ref{fig:tok-agnostic-layer}). The modular feature of the
entire framework allows a simple and generic implementation, algorithm
or functionality of each component in the control system. Moreover,
the proposed scheme is also generic for any tokamak, thus it can be
easily tested, developed and maintained. For our previous related
works, the readers can refer to \cite{Blanken_NF18} for the plasma
state and event monitoring, and to \cite{TrangVU19_FED} for the generic
actuator management strategy to deal with multiple control tasks and
actuator sharing.

The next section gives an overview of the generic PCS architecture
developed in \cite{TrangVU19_FED}. Sec. \ref{sec:Supervisor-decision}
zooms in the supervisor in the PCS, with the details of several evaluation
levels of ONE and of decision-making, as well as a concrete example
for clarification. The developed PCS is implemented on TCV and the
first results of disruption avoidance experiments are discussed in
Sec. \ref{sec:Experimental-result}. Finally Sec. \ref{sec:Conclusion}
concludes the work and gives some prospects for future works.

\section{Generic plasma control system\label{sec:Preliminary}}

\begin{figure}[t]
\includegraphics[width=3.5in]{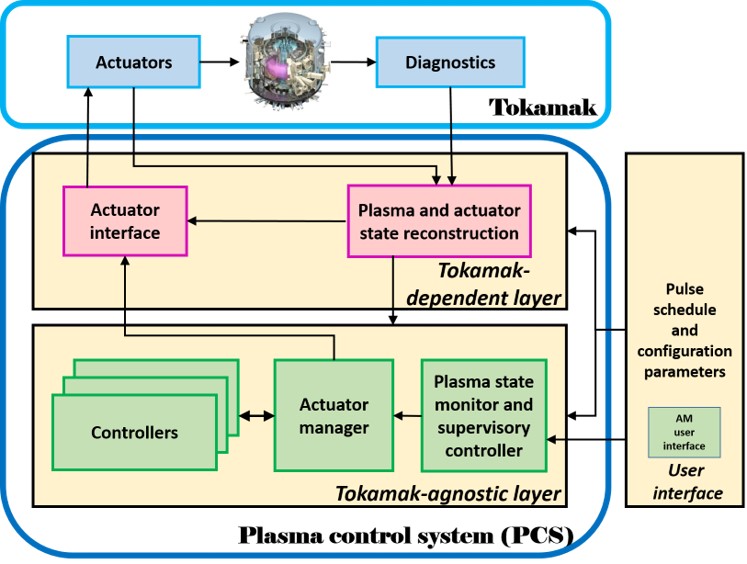}\caption{General PCS with two separated layers: tokamak-dependent and tokamak-agnostic
layers}
\label{fig:general_PCS} 
\end{figure}

The generic PCS presented in \cite{TrangVU19_FED} is revisited hereafter
with the main principles shown in Fig.  \ref{fig:general_PCS}. This
PCS is clearly separated into two layers: the tokamak- dependent layer
and the tokamak-agnostic layer. The tokamak-dependent layer includes
various real-time (RT) state reconstruction codes for plasma and actuator
states \cite{Blanken_NF18}. This layer thus converts specific plant
signals to generic continuous-value states of the plasma and actuators
which are used by the tokamak-agnostic layer and vice-versa. For example,
on TCV, the RT kinetic plasma equilibrium reconstruction can provide
realistic pressure and current density profiles \cite{Carpanese19_NF};
and the RT measures, combined with models, of heating sources provide
the states of EC and NBI actuators \cite{Poli_18,Weiland_18}. The
tokamak-agnostic layer specifically deals with the ONEs and the execution
of control tasks according to the pulse-schedule. Note that \emph{tokamak-agnostic}
is used in the sense that the functionality, algorithm and implementation
of each component are independent of the tokamak subsystems (diagnostics
and actuators), however the parameterization and specific usage are
specified by the user via the pulse schedule (user interface). Thus
the tokamak-agnostic layer can be transferable to different devices,
independently developed and maintained, while only the tokamak-dependent
layer and the user interface should be adapted for each tokamak. 

The task-based approach \cite{TrangVU19_FED} is used in the tokamak-agnostic
layer. In this approach, all decisions are made based on control tasks
and not on controllers. Generic controllers themselves cannot choose
actuators for their own interests. The controllers, on the one hand,
request actuator resources (or virtual actuators \cite{Kudlacek_FED20})
to perform their tasks, and on the other hand receive assigned generic
actuator resources per task to try to fulfill their jobs. As a result,
this scheme can avoid controller cross-talk which is the main issue
in integrated control (cf. \cite{Maljaars_FED17,TrangVU19_FED,Kudlacek_FED20}).
It also allows us to design controllers in a more generic way, independent
of the tokamak or the scenario-specific properties. Moreover, the
task-based approach greatly facilitates the interaction between the
operators and the plasma control system software, since they only
need to specify the control tasks from physics goals (or pulse schedule),
which are generic and similar among different tokamaks, regardless
of the details of the relevant controllers and actuators.

A \emph{control scenario} as pre-defined by the physicists or the
operators becomes a list of prioritized control tasks, which will
ensure the plasma evolution is as close to the target scenario as
possible. The pulse schedule is the interface between the control
scenario and a given list of tasks. Note also that the pulse-schedule
and the tunable parameters for the components in the PCS, which are
tokamak specific, are supplied via a user interface (Fig. \ref{fig:general_PCS}).

The main function of each component in the tokamak-agnostic layer
is summarized as follow:
\begin{itemize}
\item a plasma and actuator event monitor \cite{Blanken_NF18} categorizes
the state representation of the plasma, the events and the actuators
\item a supervisor evaluates the occurrence of ONEs and decides the appropriate
control scenario (list of control tasks), then activates and prioritizes
relevant tasks
\item an actuator manager defines the best actuator resource allocation
to active tasks by solving an optimization problem based on the available
actuator resources and the resource requests from controllers; and
later distributes commands to corresponding actuators
\item controllers execute control laws to fulfill their tasks with assigned
resources and also ask for new actuator resources for the next time
step
\end{itemize}
The modular and interface-standardized features of the tokamak-agnostic
layer allow us to reduce implementation errors, as well as improve
maintenance and development capabilities. For more details about the
tokamak-agnostic layer and the interfaces of each component in this
layer, the reader can refer to our previous work \cite{TrangVU19_FED}
and the references therein.

\section{Supervisor decision \label{sec:Supervisor-decision}}

We will focus on a strategy of supervisor decision to deal with ONEs
(e.g. magneto-hydrodynamic instabilities such as Neoclassical Tearing
Mode (NTM), locked mode; or the events when the plasma/actuator states
approach the physical/technical limits, such as density limit, actuator
amplitude limit, actuator energy limit, etc.) which can lead to plasma
disruption or plasma performance deterioration. Here the centralized
supervisor plays the main role to decide the action for ONE handling.
Associated with the centralized actuator resource allocator, this
supervisory level can ensure a non-conflicting and flexible use of
available actuators. A series of control scenarios are prepared in
the pulse schedule corresponding to various actions to be selected
in real-time by the supervisor.\begin{figure*}[h]

\includegraphics[width=0.9\textwidth]{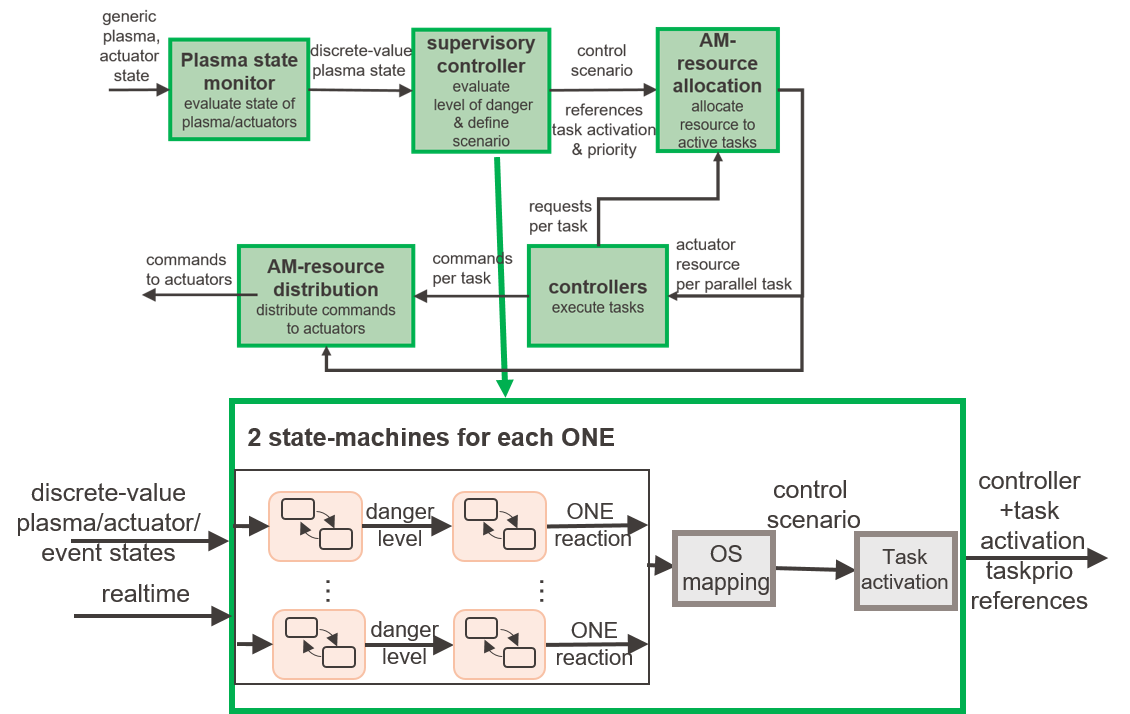}\caption{\label{fig:tok-agnostic-layer}Tokamak-agnostic
layer and zoom in on the supervisor}

\end{figure*} 

Fig. \ref{fig:tok-agnostic-layer} presents the relevant modules of
the tokamak-agnostic layer, as well as a zoom of the supervisor which
determines the control scenario output, based on the discrete-value
plasma state and actuator state inputs. For each ONE, first the \emph{danger
level} and the \emph{ONE-reaction level} are determined. Then a ONE
to Scenario (OS) mapping is used to decide the appropriate control
scenario based on a given set of ONEs and the associated \emph{ONE-reaction}
\emph{levels}.

Two finite-state-machines are used to classify the danger level and
reaction level of each ONE (Fig. \ref{fig:tok-agnostic-layer}). It
is important to notice that the thresholds for the transitions from
one state to another in the finite-state-machines are customized in
the user interface as tunable parameters, which can be modified from
shot to shot. Two lists of states of these finite-state-machines are
also shown in TABLE \ref{tab:danger_level}.

\begin{table}[h]
\caption{\label{tab:danger_level}\emph{State lists of the danger level (a),
and of the ONE reaction level (b)}}

\begin{centering}
a. \includegraphics[height=4cm]{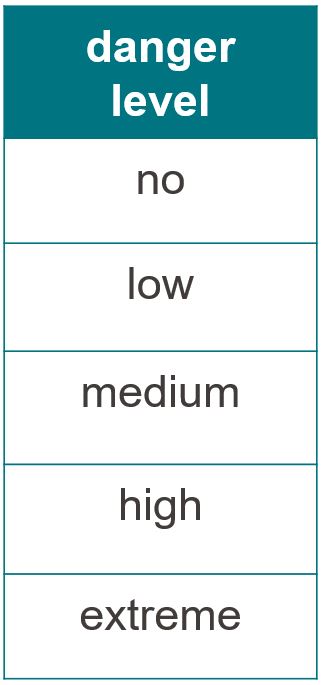} \qquad{}b.
\includegraphics[height=4cm]{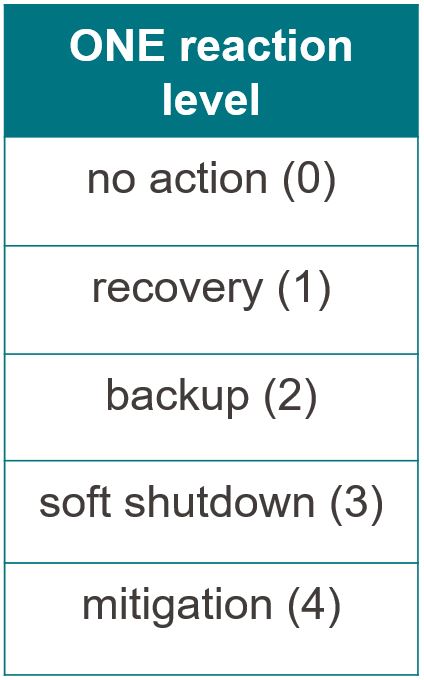}
\par\end{centering}
\centering{}The number of danger levels and reaction levels are independent
and only coincidental the same here.
\end{table}

\subsection{Danger level}

Five (states) levels of danger are defined in TABLE \ref{tab:danger_level}.a.
The classification of a ONE danger level is based either on one generic
state (amplitude state, position state, etc) or on the combination
of various generic states from the plasma event monitor. On the other
hand, in order to avoid ambiguity while several ONEs simultaneously
appear and their combination can significantly change the situation,
a virtual ONE using their combination should be created as a new independent
event. For example, a locked mode in \emph{low} danger level will
really become significant if there is also an observed increase in
radiated power. In this case, a combined event must be considered
separately from the lock mode and the radiated power events.

\subsection{Off-normal-event (ONE)-reaction level}

The \emph{danger level} is then used to define the \emph{ONE-reaction-level}
whose states are listed in TABLE \ref{tab:danger_level}.b and are
enumerated from $0$ to $4$, corresponding to the five basic types
of control scenarios that we consider at this stage (see \ref{subsec:OSmapping}).
Examples of finite-state-machines for ONE reaction are shown in Fig.
\ref{fig:Example-with-two NTM1}.a and \ref{fig:Example-with-two NTM2}.a.
It's worth noting that some states are irreversible, for instance,
state level $3$ and $4$. These states correspond to irreversible
actions such as \emph{soft-shutdown }or \emph{mitigation. }It is also
important to note that the mapping between the \emph{danger levels
}and the \emph{ONE-reaction levels }is specific for each ONE, thus
they are not always 1-1 corresponding (see e.g. Fig. \ref{fig:Example-with-two NTM1}.a,
the difference between $NTM_{21}$ and $NTM_{43}$ events).

\subsection{\label{subsec:OSmapping}ONE reaction to scenario (OS) mapping}

To select the control scenario to be executed in the current time-step,
an OS mapping (Fig. \ref{fig:Example-with-two NTM1}.b) based on the
combination of the\emph{ ONE-reaction levels} of all active ONEs is
thus necessary. A finite number of control scenarios are derived from
the given pulse-schedule. We define five basic types of control scenarios:
\emph{normal, recovery, backup, soft-shutdown, }and\emph{ disruption-mitigation}.
Several control scenarios of the same type can be defined. For example,
we often have one \emph{normal,} which is the desired/original/basic
scenario, one \emph{soft-shutdown, }and one\emph{ disruption-mitigation},
but several \emph{recovery }and several \emph{backup} scenarios depending
on the considered ONEs and the control actions on them. For instance,
with two considered ONEs, if only the first ONE needs a recovery control
action (ONE-reaction level $1$), \emph{recovery\_1 }scenario is selected;
if only the second ONEs gets ONE-reaction level $1$, \emph{recovery\_2
}scenario is chosen instead; otherwise if both of them need to be
recovered, \emph{recovery\_3 }scenario is selected this time. Finally,
a list of control tasks is determined by the user for each control
scenario in order to achieve the desired control action, as shown
in the example in TABLE \ref{tab:scenario_task_list}. 

Once the appropriate control scenario is selected based on the actual
plasma situation, the relevant control tasks will be activated, and
the corresponding references are taken into account according to the
user setting for each control scenario before the discharge. The actuator
manager and the controllers perform their normal functionality without
any judgment on the ONEs. 

\subsection{\label{subsec:Example}Example}

\begin{figure}[h]
\begin{centering}
a. \includegraphics[width=0.95\columnwidth]{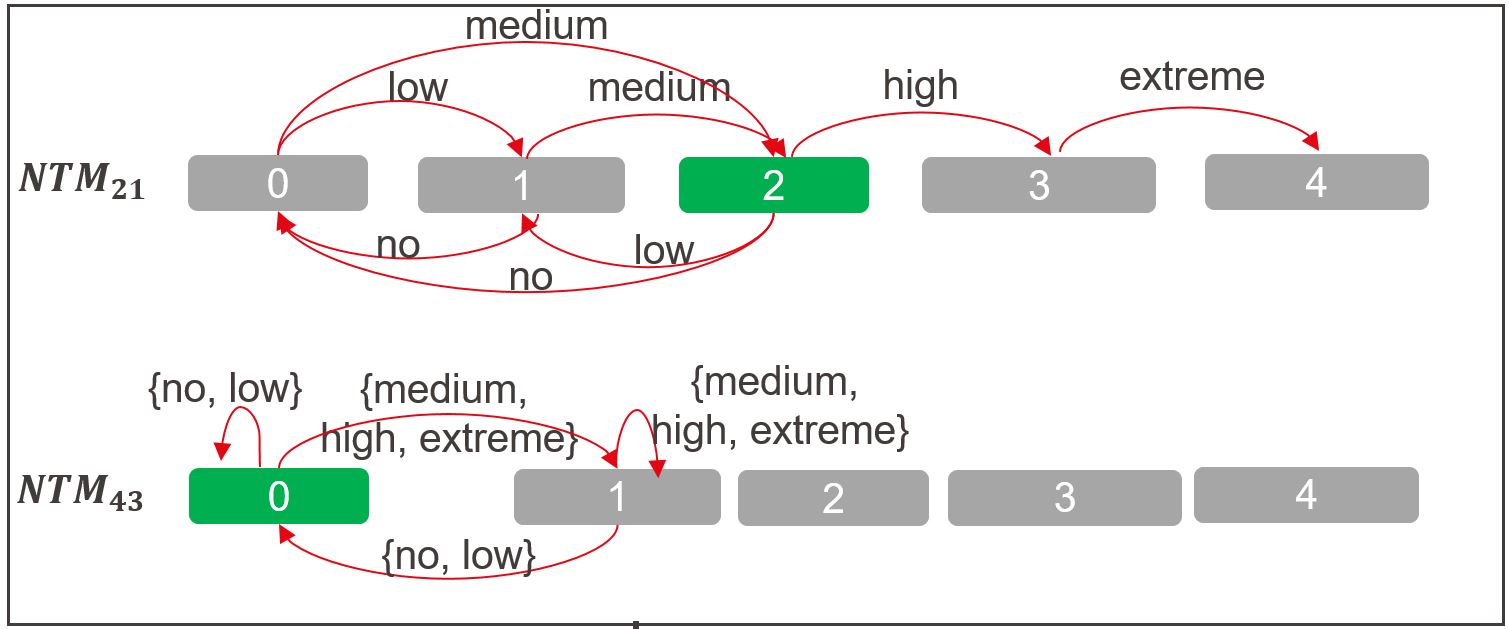}
\par\end{centering}
\begin{centering}
b. \includegraphics[width=0.95\columnwidth]{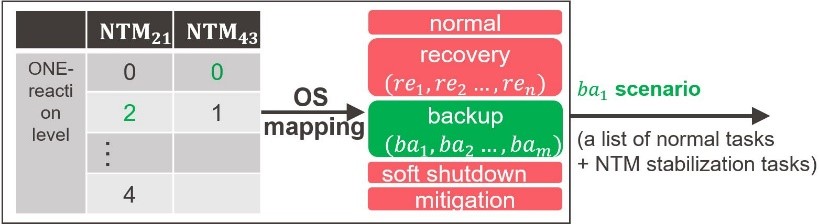}
\par\end{centering}
\caption{\label{fig:Example-with-two NTM1}Example with two NTM events in situation
1}
\end{figure}

\begin{figure}[h]
\begin{centering}
a. \includegraphics[width=0.95\columnwidth]{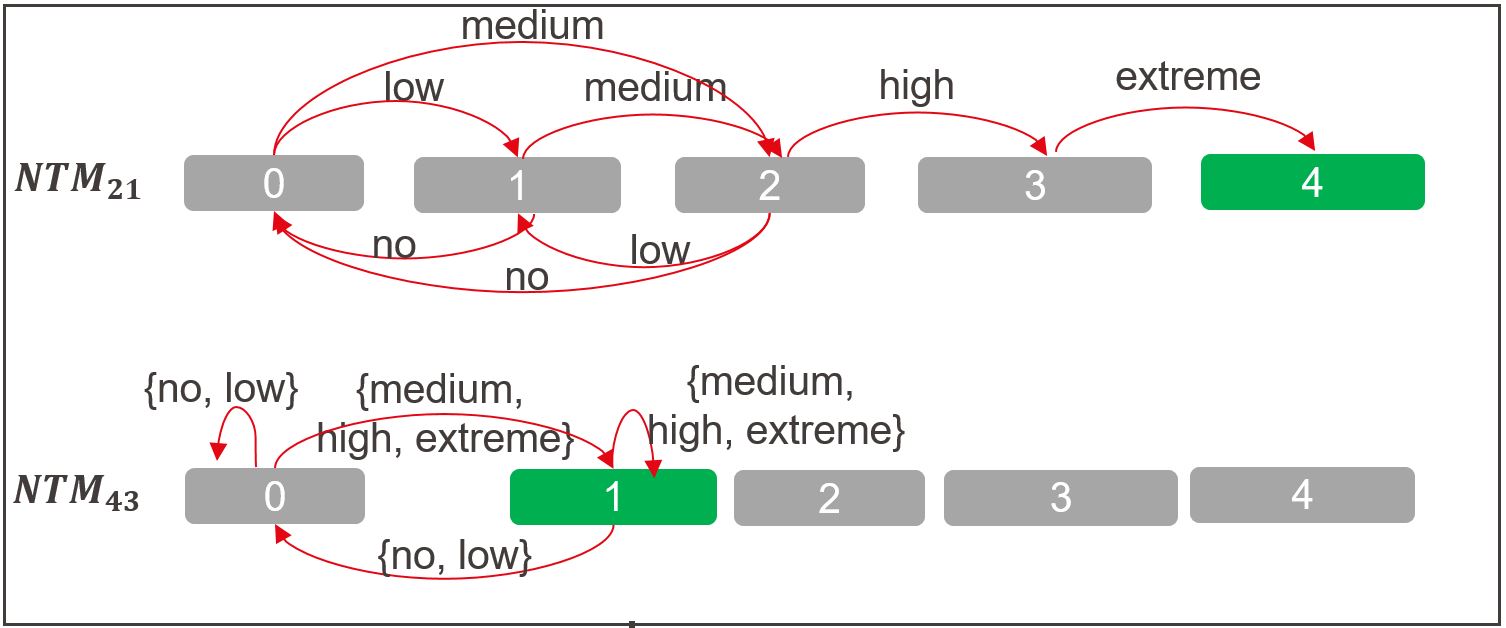}
\par\end{centering}
\begin{centering}
b. \includegraphics[width=0.95\columnwidth]{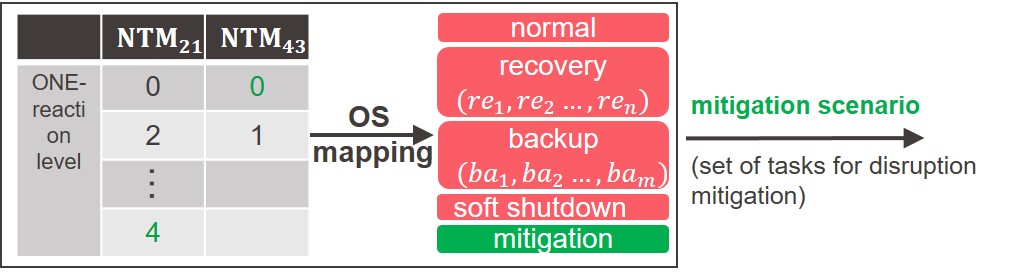}
\par\end{centering}
\caption{\label{fig:Example-with-two NTM2}Example with two NTM events in situation
2}
\end{figure}

In this subsection, two NTM events: $NTM_{21}$ and $NTM_{43}$ are
considered to be simultaneously detected. Their \emph{danger states
}are determined based on the discrete-value states of their amplitudes
from the plasma event monitor. Due to the different danger potential
of the considered events, their state-machines for the ONE \emph{reaction
level} are not the same. For instance, the $NTM_{43}$ is not very
dangerous, it only leads to a reduction of neutron productions, thus
the \emph{reaction }is either \emph{no-action }($0$) or \emph{recovery
}(\emph{$1$}). On the other hand, the $NTM_{21}$ requires all actions
up to \emph{mitigation} ($4$), since a $2/1$ mode can trigger disruptions.
Depending on the reaction level associated with these ONEs as well
as the pre-defined OS mapping, the control scenarios are different
for the two situations as shown in Fig. \ref{fig:Example-with-two NTM1}
and \ref{fig:Example-with-two NTM2}: \emph{backup1 }and \emph{mitigation
}respectively\emph{.}

In TABLE \ref{tab:scenario_task_list}, some relevant tasks are listed
associated with each control scenario in this example. A control scenario
will be chosen by the supervisor for each instant, thus the considered
tasks in this scenario will be activated based on the task activation
conditions (time intervals, event triggers, etc.). Regarding the first
situation where the \emph{backup1 }scenario is selected, three tasks
$NTM_{21}$ \emph{stabilization, $\beta$ control }and \emph{heating
feedforward }can be simultaneously activated. The actuator manager
determines the best actuator resource allocation per task based on
the actuator states as well as the requests for actuator resources
per task from the controllers. Three corresponding controllers: \emph{NTM
controller, performance controller }and \emph{feedforward controller}
are used to carry out the three considered tasks, respectively. The
controllers execute their control laws and do their best to fulfill
their given tasks with the assigned resources. Here, the \emph{NTM
controller }commands to move the EC beam deposition to the target
(\emph{$NTM$} position), and uses all EC power that it receives at
the target. The \emph{performance controller }in this case is a PID
controller which asks to modify the heating power according to the
gap between the RT estimated $\beta$ (corresponding to the total
thermal energy) and its reference. The \emph{feedforward controller
}reproduces a heating power command which is configured by the user
before the discharge. All the commands from the controllers are combined
and then sent to the tokamak-dependent layer, where they are converted
into the specific actuator commands, e.g. heating power into voltage,
radial deposition location of EC power into launcher angles, etc.

\begin{table}[h]
\caption{\label{tab:scenario_task_list}\emph{NTM event: example of tasks for
each defined control scenario}}

\begin{centering}
\includegraphics[width=3.5in]{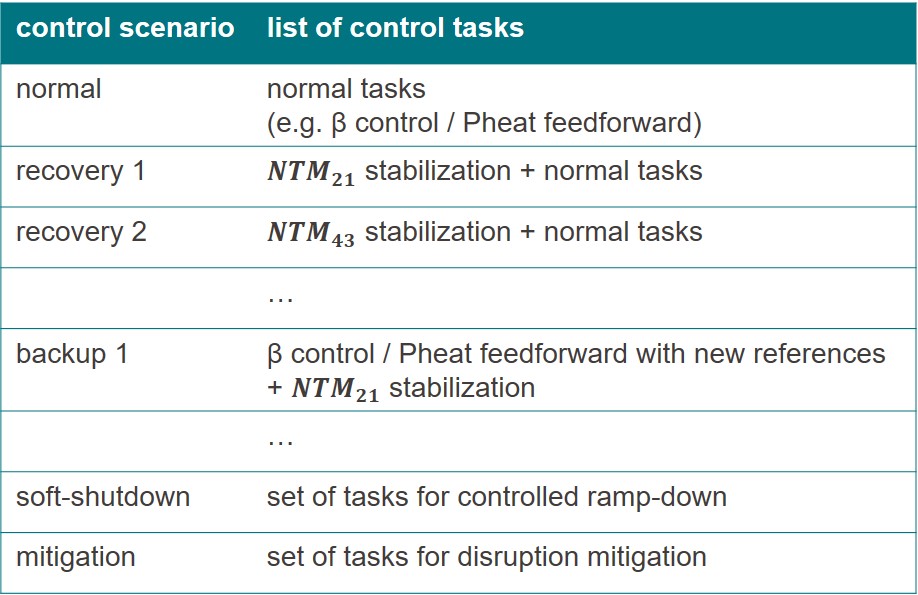}
\par\end{centering}
\centering{}Note that $\beta$ reference, feedforward reference, etc.
are tunable parameters which are given from the user interface
\end{table}

\section{Experimental result\label{sec:Experimental-result}}

\begin{figure}[h]
\begin{centering}
\includegraphics[width=0.95\columnwidth]{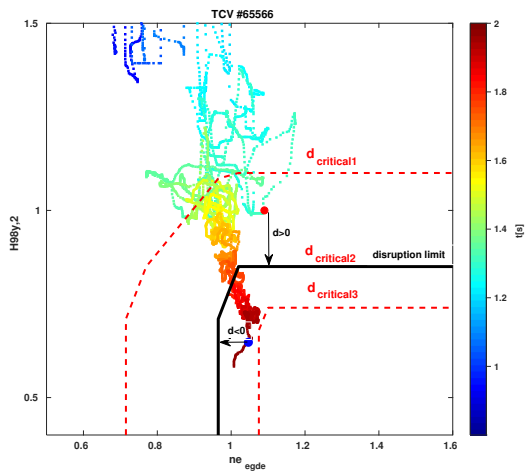}
\par\end{centering}
\caption{\label{fig:distance_65566}Density limit distance on state plane}
\end{figure}

\begin{figure}[h]
\begin{raggedright}
\includegraphics[width=1\columnwidth]{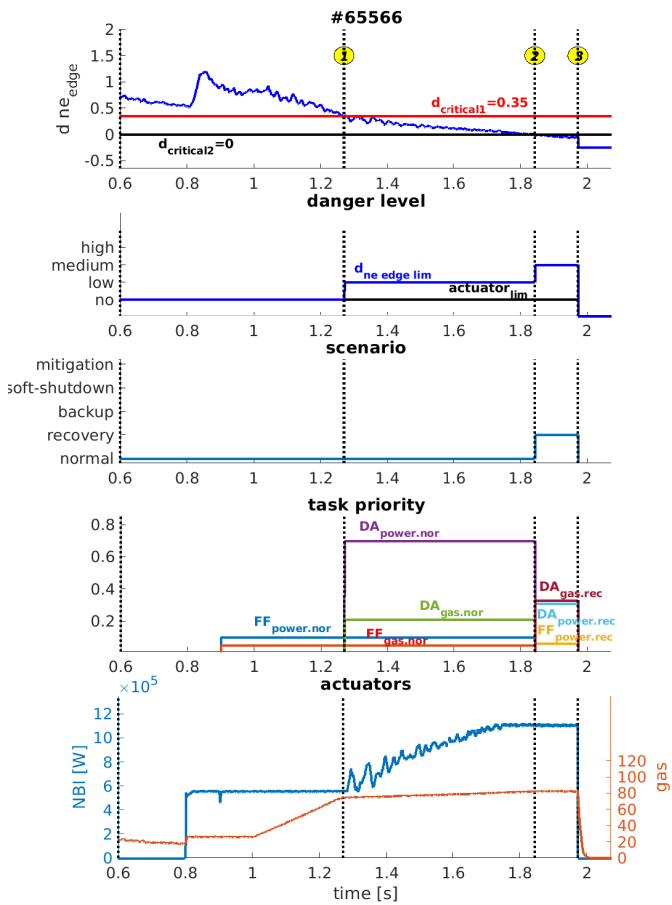}
\par\end{raggedright}
\caption{\label{fig:TCV_65566}Experimental result of TCV discharge \#$65566$:{\footnotesize{}
At {\large \textcircled{\small 1}}, $d_{ne.edge}$ violates the first
critical distance $d_{critical1}$, which activates DA tasks in the
scenario }\emph{\footnotesize{}normal}{\footnotesize{}. At {\large \textcircled{\small 2}},
$d_{ne.edge}$ goes below the second critical distance $d_{critical2}$;
the supervisor switches the scenario to }\emph{\footnotesize{}recovery
}{\footnotesize{}and activates the corresponding tasks. At {\large \textcircled{\small 3}}
the plasma finally disrupts.}}
\end{figure}

The proposed PCS has been implemented in MATLAB/ Simulink, from which
C code was generated and included in the TCV digital real-time control
system (cf. \cite{Galperti_TNS17,Felici_FED14}). An application for
disruption avoidance experiments is presented in Fig. \ref{fig:TCV_65566},
where the discharge is pushed towards the H-mode density limit disruption
by a gas-flux ramp. The main purpose of disruption avoidance experiments
is to avoid the abrupt loss of energy confinement, or even to recover
the plasma to the previous stable states. However, in this experiment,
we aim to control the plasma to slowly approach the density limit
for detailed physics studies. Several new modules have been implemented
in the TCV PCS to determine in RT: the factor $H_{98y2}$ characterizing
the energy confinement time, and the normalized edge density $ne_{edge\,norm}$
\cite{Bernert15}; together, we derive the distance $d_{ne\,edge}$
between the system states $\left(H_{98y2},ne_{edge\,norm}\right)$
and the empirical disruption limit (see Fig. \ref{fig:distance_65566}).
This distance is the key factor used by the supervisory layer to determine
an appropriate control scenario. In this example, two ONEs are considered:
$d_{ne\,edge}$ and $actuator_{lim}$ for the NBI energy limit. The
supervisor evaluates the dangers from these ONEs to switch between
different scenarios: \emph{normal, recovery}, and \emph{soft-shutdown},
in which three sets of relevant control tasks are configured beforehand. 

In Fig. \ref{fig:TCV_65566}, second panel, the danger level of the
{\footnotesize{}$d_{ne.edge}$} is \emph{low} from {\large \textcircled{\small 1}}
when the distance is below the first critical threshold $d_{critical1}$,
and is \emph{medium} from {\large \textcircled{\small 2}} when the
distance is below the second threshold $d_{critical2}$; while the
$actuator_{lim}$ danger is \emph{high} when the NBI energy reaches
the threshold $95\%$ of the total energy of $1.3MJ$ (which is not
the case in this experiment) and is \emph{no} otherwise. Here, the
customized ONE reaction levels based on the danger level of each ONE
are specified in TABLE \ref{tab: danger_reaction}. 
\begin{center}
\begin{table}[h]
\caption{\label{tab: danger_reaction}\emph{Reaction levels} \emph{corresponding
to each danger level of each ONE and the selected scenario}}

\centering{}\includegraphics[width=3.5in]{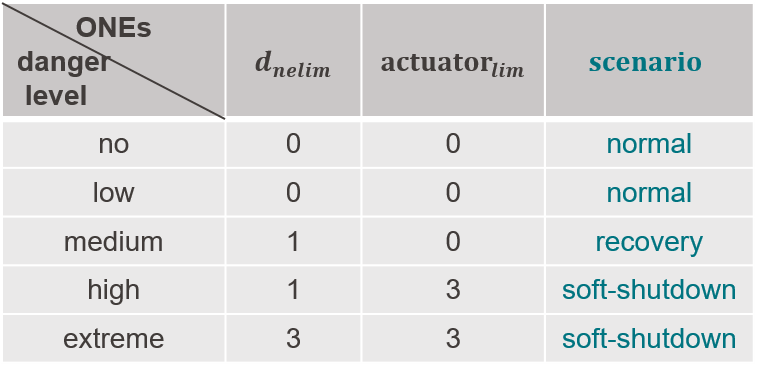}
\end{table}
\par\end{center}

The control scenario (third panel) is based on the combination of
the reaction levels of the two ONEs, which is \emph{recovery }if the
highest danger level reaches \emph{medium} from {\large \textcircled{\small 2}},
otherwise it remains \emph{normal}. According to the chosen scenario,
different control tasks are activated and prioritized (fourth panel).
In \emph{normal} scenario, the feedforward task $FF_{power.nor}$
asks for a constant heating power of $0.65MW$ and the $FF_{gas.nor}$
commands a fast gas-flux ramp. The disruption-avoidance task $DA_{power.nor}$
modifies the power according to the distance{\footnotesize{} $d_{ne.edge}$}
only when it is below the $d_{critical1}$, and the $DA_{gas.nor}$
reduces the gas-flux ramp. In \emph{recovery} scenario, the $FF_{power.rec}$
is the same as the $FF_{power.nor}$, while the $DA_{power.rec}$
asks for the maximum power and the $DA_{gas.rec}$ keeps the gas flux
constant. Consequently, on the one hand, the NBI power (last panel)
is composed of a constant power $0.65MW$ ($FF_{power.nor}$) and
an extra power ($DA_{power.nor}$) increasing proportionally up to
the maximum heating power $1.3MW$. On the other hand, the gas flux
is first increasing fast ($FF_{gas.nor}$), then slowly ($DA_{gas.nor}$)
at {\large \textcircled{\small 1}}, and finally is frozen ($DA_{gas.rec}$)
at {\large \textcircled{\small 2}}. This allows a slow and well-controlled
approach to the density limit for detailed physics studies. A \emph{soft-shutdown}
scenario is also set up with two feedforward tasks to cut off both
the NBI power and the gas flux in a controlled manner when the distance
{\footnotesize{}$d_{ne.edge}$} goes below the $d_{critical3}$. Unfortunately,
in that case, the discharge disrupted before that situation happens.

\section{Conclusion\label{sec:Conclusion}}

This work presents a supervisory strategy to deal with off-normal-events.
Each ONE is evaluated for its danger level and the necessary reaction,
then a global decision is made to define an appropriate control scenario.
The presented control architecture has been successfully implemented
and tested on the TCV tokamak in the context of density limit disruption
avoidance experiments; employing the exposed control scenario switching
methodology. It has been fruitfully capable of smoothly reaching density
limits in various plasma discharges. This architecture will be used
for different integrated control objectives such as simultaneous controls
of L-H mode, NTM, $\beta$, and $q-$profile, etc. in the upcoming
experiments.

\section*{Acknowledgment}

This work was supported in part by the Swiss National Science Foundation.
This work has been carried out within the framework of the EUROfusion
Consortium and has received funding from the Euratom research and
training programme 2014 - 2018 and 2019 - 2020 under grant agreement
No 633053. The views and opinions expressed herein do not necessarily
reflect those of the European Commission.

\bibliographystyle{unsrt}
\bibliography{SUP_bib}

\end{document}